\documentclass[twocolumn,aps,prd]{revtex4}
 \usepackage{graphicx,amssymb}
 \usepackage[utf8]{inputenc} 
 
\usepackage[urlcolor=blue,colorlinks,breaklinks=true]{hyperref}

\PassOptionsToPackage{obeyspaces,spaces,hyphens}{url}
 
 \usepackage{amsmath}

\begin{document}
 	\def\half{{1\over2}}
 	\def\shalf{\textstyle{{1\over2}}}
 	
 	\newcommand\lsim{\mathrel{\rlap{\lower4pt\hbox{\hskip1pt$\sim$}}
 			\raise1pt\hbox{$<$}}}
 	\newcommand\gsim{\mathrel{\rlap{\lower4pt\hbox{\hskip1pt$\sim$}}
 			\raise1pt\hbox{$>$}}}

\newcommand{\be}{\begin{equation}}
\newcommand{\ee}{\end{equation}}
\newcommand{\bq}{\begin{eqnarray}}
\newcommand{\eq}{\end{eqnarray}}
 	

\title{Boltzmann's $H$-theorem, entropy and the strength of gravity in theories with a nonminimal coupling between matter and geometry}

\author{P.P. Avelino}
\email[Electronic address: ]{pedro.avelino@astro.up.pt}
\affiliation{Instituto de Astrof\'{\i}sica e Ci\^encias do Espa{\c c}o, Universidade do Porto, CAUP, Rua das Estrelas, PT4150-762 Porto, Portugal}
\affiliation{Departamento de F\'{\i}sica e Astronomia, Faculdade de Ci\^encias, Universidade do Porto, Rua do Campo Alegre 687, PT4169-007 Porto, Portugal}
\affiliation{School of Physics and Astronomy, University of Birmingham, Birmingham, B15 2TT, United Kingdom}

\author{R.P.L. Azevedo}
\email[Electronic address: ]{rplazevedo@fc.up.pt}
\affiliation{Instituto de Astrof\'{\i}sica e Ci\^encias do Espa{\c c}o, Universidade do Porto, CAUP, Rua das Estrelas, PT4150-762 Porto, Portugal}
\affiliation{Departamento de F\'{\i}sica e Astronomia, Faculdade de Ci\^encias, Universidade do Porto, Rua do Campo Alegre 687, PT4169-007 Porto, Portugal}
 	
\date{\today}
\begin{abstract}
In this paper we demonstrate that Boltzmann's $H$-theorem does not necessarily hold in the context of theories of gravity with nonminimally coupled matter fields. We find sufficient conditions for the violation of Boltzmann's $H$-theorem and derive an expression for the evolution of Boltzmann's $H$ in terms of the nonminimal coupling function, valid in the case of a collisionless gas in a homogeneous and isotropic Friedmann-Lema\^itre-Robertson-Walker universe. We highlight the implications of this result for the evolution of the entropy of the matter fields, briefly discussing the role played by collisions between particles whenever they are relevant. We also suggest a possible link between the high entropy of the Universe and the weakness of gravity in the context of these theories.
\end{abstract}

\maketitle
 	
\section{Introduction}
\label{sec:intr}

General relativity has been extremely successful in describing the gravitational interaction from cosmological scales down to the sub-millimetre scale \cite{Will:2014kxa}. However, this success comes at the price of having to postulate an exotic dark energy component (dubbed dark energy), violating the weak energy condition, in order to explain the observed acceleration of the Universe \cite{Riess:1998cb,Perlmutter:1998np,Aghanim:2018eyx}. A similar component, albeit with a much larger energy density, is also required for primordial inflation to occur, providing a solution to some of the most fundamental problems of the standard cosmological model (see, for example,  \cite{Akrami:2018odb} and references therein). 

The search for extensions of general relativity which can more naturally explain the early and late dynamics of the Universe is an extremely active area of research. Broad classes of modified theories of gravity incorporate the possibility of a nonminimal coupling (NMC) between geometry and matter \cite{Nojiri:2004bi,BouhmadiLopez:2004ys,Allemandi:2005qs,Bertolami:2007gv,Sotiriou:2008it,Bertolami:2008ab,Harko:2010mv,Harko:2011kv,Harko:2012hm}. In these theories the energy-momentum tensor is not usually covariantly conserved, which results in the non-geodesic motion of point particles (the same applies to solitonic particles with fixed rest mass and structure, independently of the details of their composition \cite{Avelino:2018qgt,Avelino:2018rsb}). The resulting forces are dependent on the individual linear momentum of the particles, and provide an extra contribution to their linear momentum evolution. This contribution is not only responsible for the tight cosmic microwave background and primordial  nucleosynthesis constraints on the strength of the NMC coupling between the gravitational and matter fields \cite{Azevedo:2018nvi}, but can also have profound implications for the evolution of the Universe, particularly at early times, which we shall explore in the present paper.

In the late nineteenth century Boltzmann developed, almost single-handedly, the foundations of  modern statistical mechanics. One of his major contributions, Boltzmann's $H$-theorem, implies that, under rather general conditions, the entropy of a closed system is a non-decreasing function of time \cite{1965AmJPh..33..391J}. In a recent work \cite{Bertolami:2020ldj} it has been claimed that Boltzmann's $H$-theorem is preserved in the context of NMC theories of gravity. In this paper we shall demonstrate that this is not necessarily the case by determining the evolution of the phase-space particle distribution function under the momentum-dependent forces associated to the NMC to gravity. We shall also investigate the impact of a NMC between the gravitational and matter fields on the evolution of the entropy and energy densities of the early Universe, suggesting a possible link between the high entropy of the Universe and the weakness of gravity.


The outline of this paper is as follows. In Sec. \ref{sec:model} we briefly describe the simplest possible theory allowing for a NMC coupling between the gravitational and the matter fields. In Sec. \ref{sec:force} we compute  the momentum-dependent forces on point-particles associated with the NMC to gravity. In Sec. \ref{sec:boltzmann} we  consider the phase space continuity equation in the context of NMC gravity in the absence of particle collisions and study its implications for the evolution of Boltzmann's $H$ in a homogeneous and isotropic Friedmann-Lema\^itre-Robertson-Walker (FLRW) background. We also provide sufficient conditions for the violation of Boltzmann's $H$-theorem. In Sec. \ref{sec:entropy} we consider the evolution of the entropy of the universe in NMC gravity, exploiting the relationship between Boltzmann's $H$-theorem and the second law of thermodynamics, briefly discussing the  impact of particle collisions. The connection between the evolution of the entropy of the Universe and the strength of gravity is also explored in Sec. \ref{sec:entropy}. Finally, we conclude in section \ref{sec:conc}. 

Throughout this paper we use units such that $16\pi G=c=k_B=h=1$, where $G$ is Newton's  gravitational constant, $c$ is the value of the speed of light in vacuum, $k_B$ is the Boltzmann  constant and $h$ is the Planck constant. We also adopt the metric signature $(-,+,+,+)$. The Einstein summation convention will be used whenever a Greek or a Latin index variable appears twice in a single term, once in an upper (superscript) position and once in a lower (subscript) position. Greek and Latin indices take the values $0,\cdots,3$ and $1,...,3$, respectively.

\section{Nonminimally coupled gravity}
\label{sec:model}

Consider a theory with a NMC between gravity and matter described by the action
\begin{equation}
\label{eq:action}
S=\int d^4 x \sqrt{-g} \left[R+ \mathcal F(R)\mathcal{L}_{\rm m}\right]\,,
\end{equation}
where $g$ is the determinant of the metric, $\mathcal{L}_{\rm m}$ is the Lagrangian of the matter fields, and $\mathcal F > 0$ is a function of the Ricci scalar $R$. Note that general relativity is recovered if $\mathcal F(R)=1$. Also, the generalization from $\mathcal L = R+ \mathcal F(R)\mathcal{L}_{\rm m}$ to  $\mathcal L = \mathcal F_1(R)+ \mathcal F_2(R)\mathcal{L}_{\rm m}$ is straightforward --- the former is considered in the present paper for simplicity (this choice does not affect our main results). Assuming a Levi-Civita connection, the equations of motion for the gravitational field are given by
\begin{eqnarray}
\label{eq:eqmotion}
(1  + \mathcal F'\mathcal{L}_{\rm m}) G^{\mu\nu}&=&\half  \mathcal F \, T^{\mu\nu}+\Delta^{\mu\nu}(\mathcal F'\mathcal{L}_{\rm m})\nonumber\\
&-&\half R \mathcal F'\mathcal{L}_{\rm m} g^{\mu\nu}\, ,
\end{eqnarray}
where $g_{\mu\nu}$ are the components of the metric, $G^{\mu\nu}=R^{\mu\nu}-\shalf g^{\mu\nu} R $ and $R^{\mu\nu}$ are, respectively, the components of the Einstein and Ricci tensors, $\Delta^{\mu \nu} \equiv \nabla^\mu \nabla^\nu - g^{\mu \nu} \Box$, $\Box \equiv \nabla^\mu \nabla_\mu$, $\nabla_\mu$ is the covariant derivative with respect to the space-time coordinates $x^\mu$ and a prime represents a derivative with respect to $R$. The energy-momentum tensor of the matter fields may be computed as
\begin{equation}
\label{eq:energymom}
T^{\mu\nu}={2\over \sqrt{-g}}{\delta(\sqrt{-g}\mathcal{L}_{\rm m})\over \delta g_{\mu\nu}}\,.
\end{equation}

Taking the covariant derivative of Eq.~\eqref{eq:eqmotion} and using the Bianchi identities one obtains
\begin{equation}
\label{eq:noncons}
\nabla_\nu T^{\mu\nu}=\frac{\mathcal F'}{\mathcal F}(g^{\mu\nu}\mathcal{L}_{\rm m}-T^{\mu\nu})\nabla_\nu R\,.
\end{equation}
Equation~\eqref{eq:noncons} implies that the energy-momentum tensor is, in general, not covariantly conserved. In the following section we shall discuss the implications of this result for point particles.

\section{4-force on point particles}
\label{sec:force}

In this section we shall compute the 4-force on a point particle associated to the NMC to gravity. Consider the action 
\begin{equation}
\label{eq:actionpp}
S=-\int d \tau \, m 
\end{equation}
of a single point particle with energy-momentum tensor
\begin{equation}
T^{\mu \nu} = \frac{m}{\sqrt {-g}}\int d \tau \, u^\mu u^\nu \delta^4(x^\sigma-\xi^\sigma(\tau)),
\end{equation}
where $\delta^4(x^\sigma-\xi^\sigma(\tau))$ denotes the four–dimensional Dirac delta function, $\xi^\sigma(\tau)$ represents the particle wordline, $\tau$ is the proper time, $u^\mu$ are the components of the particle 4-velocity ($u^\mu u_\mu=-1$), and $m$ is the proper particle mass. If one considers its trace $T=T^{\mu \nu} g_{\mu \nu}$ and integrates over the whole of space-time, one  obtains
\begin{eqnarray}
\int d^{4}x \sqrt{-g} \, T &=&- \int d^4x \,d\tau\, m\, \delta^4\left(x^\sigma-\xi^\sigma(\tau)\right) \nonumber\\
&=&- \int d\tau \,m \,.
\end{eqnarray}
This can immediately be identified as the action for a single massive particle  and, therefore, it implies that
\begin{equation}
\label{eq:lag}
{\mathcal L}_{\rm m} = T= -  \frac{m}{\sqrt {-g}}\int d \tau \, \delta^4(x^\sigma-\xi^\sigma(\tau))
\end{equation}
is the particle Lagrangian. Note that the particle Lagrangian given in Eq.~\eqref{eq:lag} may be obtained from the action for a solitonic particle, with fixed rest mass and structure, and negligible  self-induced gravitational field, independently of its constitution \cite{Avelino:2018rsb} (see also \cite{Polyakov:2018zvc,Avelino:2019esh}). The covariant derivative of the energy-momentum tensor may be written as
\begin{equation}
\nabla_\nu T^{\mu \nu} =\frac{m}{\sqrt {-g}}\int d \tau \left(\nabla_\nu u^\mu\right) u^\nu \delta^4(x^\sigma-\xi^\sigma(\tau))\,.
\end{equation}
Hence, Eq.~\eqref{eq:noncons} becomes
\begin{eqnarray}
0&=& \frac{m}{\sqrt {-g}}\int d \tau  \delta^4(x^\sigma-\xi^\sigma(\tau)) \nonumber\\
&\times& \left(\frac{d u^\mu}{d \tau} +\Gamma^\mu_{\alpha \beta} u^\alpha u^\beta+ \frac{\mathcal F'}{\mathcal F}  h^{\mu \nu} \nabla_\nu R \right)\,,
\end{eqnarray}
where $h^{\mu \nu}=g^{\mu \nu}+ u^\mu u^\nu$ is the projection operator. The equation of motion of a point particle is then given by
\begin{equation}
\label{eq:accelerationf0}
\mathfrak{a}^\mu=\frac{d u^\mu}{d \tau} +\Gamma^\mu_{\alpha \beta} u^\alpha u^\beta=-
 \frac{\mathcal F'}{\mathcal F}   h^{\mu \nu} \nabla_\nu R \,,
\end{equation}
where
\begin{equation}
\label{eq:force}
\mathfrak{f}^{\nu}=m \mathfrak{a}^\mu=-m \frac{\mathcal F'}{\mathcal F}   h^{\mu \nu} \nabla_\nu R 
\end{equation}
is the momentum-dependent 4-force on a particle associated to the NMC to gravity and $\mathfrak{a}^\mu$ is the corresponding 4-acceleration (see \cite{Ayaita:2011ay} for an analogous calculation in the context of growing neutrino models where the neutrino mass is non-minimally coupled to a dark energy scalar field). 

\subsection{A note on the 4-acceleration of a fluid element}

If the particles are part of a fluid, then the 4-acceleration of the individual particles does not, in general, coincide with the 4-acceleration of the fluid element to which they belong. This may be shown explicitly by considering a perfect fluid with energy-momentum tensor
\begin{equation}
\label{eq:pfemt}
T^{\mu\nu}=(\rho+p)U^\mu U^\nu + p g^{\mu\nu}\,,
\end{equation}
where $\rho$ and $p$ are respectively the proper energy density and pressure of the perfect fluid and $U^\mu$ are the components of its 4-velocity. The 4-acceleration equation of a perfect fluid element is given by \cite{Bertolami:2007gv,Bertolami:2008ab}
\begin{eqnarray}
\label{eq:accelerationf}
& &\mathfrak{a}^\mu_{[\rm fluid]}=\frac{dU^{\mu}}{d\tau}+\Gamma^\mu_{\alpha\beta} U^\alpha U^\beta \nonumber \\
&=&\frac{1}{\rho+p}\left[\frac{ \mathcal F'}{\mathcal F} (\mathcal{L}_{\rm m[fluid]}-p)\nabla_\nu R - \nabla_\nu p\right]h^{\mu\nu}_{[\rm fluid]}\,,
\end{eqnarray}
where ${\mathcal L}_{\rm m[fluid]}$ is the Lagragian of the perfect fluid and $h^{\mu \nu}_{[\rm fluid]}=g^{\mu \nu}+ U^\mu U^\nu$ is the corresponding projection operator. Hence, not only the 4-velocity of the fluid $\bf U$ is in general very different from the velocity 4-velocity $\bf u$ of the individual particles at any point on the fluid, but the same is also true in the case of the 4-acceleration (this point has been overlooked in \cite{Bertolami:2020ldj}).

Moreover, Eq.~\eqref{eq:accelerationf} shows that the 4-acceleration of a perfect fluid element has an explicit dependence on its on-shell Lagrangian. The appropriate on-shell Lagrangian in the case of an ideal gas is $\mathcal L_{\rm m[ideal\ gas]}=T=-\rho+3p$ \cite{Avelino:2018qgt,Avelino:2018rsb} --- see \cite{Ferreira:2020fma} for a detailed discussion of the appropriateness of the use of different Lagrangians to describe various components of the cosmic energy budget, including a discussion of the reasons why the on-shell Lagrangians $\mathcal L_{\rm m[fluid]}=p$ and $\mathcal L_{\rm m[fluid]}=-\rho$ of Brown's work on the action functionals for relativistic perfect fluids \cite{Brown:1992kc} (see also Schutz \cite{SCHUTZ19771}) should not in general be used in the case of an ideal gas with a NMC to gravity, the exception being $\mathcal L_{\rm m[dust]}=-\rho$. In fact, in the case of dust ($p=0$) the 4-velocity is the same for all the particles, thus implying that the 4-acceleration of the fluid is equal to the 4-acceleration the particles. Hence, by considering the 4-acceleration of dust one may again recover the result given in Eq.~\eqref{eq:accelerationf0} for the 4-acceleration of the individual particles:
\begin{eqnarray}
\label{eq:accelerationfnew}
\mathfrak{a}^\mu&=&\mathfrak{a}^\mu_{[\rm dust]}=\frac{\mathcal{L}_{\rm m[ dust]}}{\rho}\frac{ \mathcal F'}{F} h^{\mu\nu}_{[\rm dust]} \nabla_\nu R \nonumber \\ 
&=& -\frac{ \mathcal F'}{F} h^{\mu\nu} \nabla_\nu R \,.
\end{eqnarray}

In the case of a photon gas (with $p=\rho/3$), the appropriate on-shell Lagrangian in Eq.~\eqref{eq:accelerationf} would be identically equal to zero ($\mathcal L_{\rm m[photon \  gas]}=T=-\rho+3p=0$). This is not surprising since the electromagnetic Lagrangian,
\begin{equation} 
{\mathcal L}_{\rm EM}=-\frac{1}{4}F^{\mu\nu}F_{\mu\nu}\,,
\end{equation} 
vanishes on-shell for electromagnetic waves in vacuum.

\section{Boltzmann's H-theorem}
\label{sec:boltzmann}

The standard collisionless Boltzmann's equation, given in Minkowski space by
\begin{equation}
\label{eq:boltzmann1}
\frac{d f}{dt}= \frac{\partial  f}{\partial t}+ \frac{d \vec r}{dt} \cdot \nabla_{\vec r} f +  {\vec F} \cdot \nabla_{\vec p} f=0 \,,
\end{equation}
expresses the constancy in time of a six-dimensional phase space volume element ${\mathcal V}_6$ containing a fixed set of particles in the absence of particle collisions. Here, $t$ is the physical time, the six-dimensional phase space is composed of the six positions and momentum coordinates $({\vec r},{\vec p})$ of the particles, ${\vec F}=d{\vec p}/dt$ is the 3-force on the particles (assumed to be independent of ${\vec p}$), and $f(t,{\vec r},{\vec p}) {\mathcal V}_6 $ is the number of particles in the six-dimensional infinitesimal phase space volume element ${\mathcal V}_6 = d^3 r \, d^3 p$. However, in the presence of a NMC to gravity $\vec F$ can depend on $\vec p$ and, therefore, ${\mathcal V}_6$ is in general no longer conserved. In this case, phase-space continuity, expressing particle number conservation in six-dimensional phase space in the absence of collisions,
 \begin{equation}
 \label{eq:continuity}
\frac{\partial  f}{\partial t}+\nabla_{\vec r} \cdot \left(f \frac{d \vec r}{dt}\right)  + \nabla_{\vec p} \cdot \left(f \vec F \right)=0 \,,
 \end{equation}
should be used rather than Eq.~\eqref{eq:boltzmann1}.
Here, $\vec r$ and $\vec p$ are independent variables, thus implying $\nabla_{\vec r} \cdot {\vec p}=0$. Note that no assumption has been made regarding the relativistic or non-relativistic nature of the particles (Eq.~\eqref{eq:continuity} applies in both regimes).

In a flat homogeneous and isotropic universe, described by the FLRW metric, the line element is given by
\begin{equation}
ds^2=-dt^2+d{\vec r} \cdot d {\vec r}= -dt^2 + a^2(t) d{\vec q} \cdot d{\vec q}\,,
\end{equation}
where $a(t)$ is the scale factor and ${\vec q}$ are comoving cartesian coordinates. In this case, the Ricci scalar is a function of cosmic time alone [$R=R(t)$] and the $i0$ components of the projection operator may be written as  $h^{i0}=\gamma^2 v^i$, where $\gamma=u^0=dt/d\tau$ and $v^i=u^i/\gamma$ are the components of the 3-velocity. Therefore, Eq.~\eqref{eq:force} implies that the momentum-dependent 3-force on the particles is given by
\begin{eqnarray}
\label{eq:3-force}
F^i=\frac{d {p}^i}{dt}&=& \frac{\mathfrak{f}^{i}}{\gamma} -\frac{d\ln a}{dt}p^i= -\left(\frac{d\ln a}{dt}+\frac{\mathcal F'}{\mathcal F}   \frac{d R} {dt} \right)p^i \nonumber \\
&=&- \left(\frac{d \ln a}{dt}+ \frac{d \ln \mathcal F}{dt}   \right) p^i  \nonumber\\
&=&  -\frac{d \ln \left(a \mathcal F \right)}{dt} p^i \,,
\end{eqnarray}
This in turn implies that $p^i \propto (\mathcal F a)^{-1}$, so that 
\begin{equation}
\label{V6}
{\mathcal V}_6 = d^3 r \, d^3 p \propto   {\mathcal F}^{-3}\,,
\end{equation}
where we have taken into account that $d^3 r= a^3 d^3 q$. Eq.~\eqref{V6} explicitly shows that in the presence of a NMC to gravity the phase space volume is, in general, no longer incompressible.

In a homogeneous and isotropic universe
\bq
\frac{d \vec r}{dt}&=& \frac{da}{dt} \vec q + a \frac{d \vec q}{dt} =  \frac{d \ln a}{dt}\vec r + \vec v \nonumber\\
&=&  \frac{d \ln a}{dt}\vec r + \frac{\vec p}{(m^2 +p^2)^{1/2}}\,,
\eq
where $m$ is the rest mass of the particles, thus implying that
\be
\label{drdt}
\nabla_{\vec r} \cdot \left( \frac{d \vec r}{dt} \right) =3 \frac{d \ln a}{dt} \,.
\ee

Substituting Eqs.~\eqref{eq:3-force} and~\eqref{drdt} into the phase space continuity equation --- note that  Eq.~\eqref{eq:continuity} remains valid in a FLRW background --- and taking into account that in a homogeneous universe  $f$ is independent of ${\vec r}$ [$f=f(t,{\vec p})$], one obtains
\begin{eqnarray}
\label{eq:boltzmann2}
0 &=& \frac{\partial  f}{\partial t} +  f \nabla_{\vec r} \cdot \left( \frac{d \vec r}{dt} \right)+ \vec F  \cdot  \nabla_{\vec p} f +  f \nabla_{\vec p} \cdot {\vec F}\nonumber \\ 
&=&\frac{\partial  f}{\partial t} - \frac{\partial  f}{\partial p^i}\frac{d \ln \left(a \mathcal F\right)}{dt} p^i -3 f \frac{d \ln \mathcal F}{dt}  \,.
\end{eqnarray}
Note that Eq.~\eqref{eq:boltzmann2} does not include colision terms and, therefore, it only applies in the case of collisionless fluids. For example, after neutrino decoupling non-gravitational neutrino interactions may in general be neglected and, consequently, Eq.~\eqref{eq:boltzmann2} may be used to determine the evolution of the neutrino phase space distribution function for as long as the Universe remains approximately homogeneous and isotropic (the same applying to photons after recombination, although to a lesser extent). We shall defer to the  following section a discussion of the impact of collisions in situations where they might be relevant.

Let us start by explicitly verifying the conservation of the number of particles $N$ inside a constant comoving spatial volume $V_q$ defined by $\int d^3 r =a^3 \int d^3 q =a^3 V_q$ ($N=\int d^3 r \, d^3 p \, f=a^3 V_q \int f d^3 p$)
\begin{eqnarray}
\label{eq:dotN}
\frac{d N}{dt}&=& 3 \frac{d\ln a}{dt} N+a^3 V_q \int  d^3 p \frac{\partial f}{\partial t}  \nonumber\\
&=& 3 \frac{d\ln (a \mathcal F)}{dt} N\nonumber  \\
&+& a^3 V_q \int  d^3 p \frac{\partial  f}{\partial p^i}\frac{d \ln \left(a \mathcal F\right)}{dt} p^i  = 0\,.
\end{eqnarray}
Here, we have used Eq.~\eqref{eq:boltzmann2} in order to evaluate $\partial f/ \partial t$ and performed the momentum integral by parts.

Let us now consider Boltzmann's $H$ defined by
\begin{equation}
H= \int d^3r \, d^3 p f \ln f = a^3 V_q \int d^3 p f \ln f  \,,
\end{equation}
Taking the derivative of $H$ with respect to the physical time and using Eq.~\eqref{eq:dotN} one obtains
\begin{eqnarray}
\label{eq:dotH}
\frac{d H}{dt}&=& 3 \frac{d\ln a}{dt} H+a^3 V_q \int  \, d^3 p  (1+\ln f)  \frac{\partial f}{\partial t} \nonumber\\
&=& 3 \frac{d\ln a}{dt} (H-N)+ a^3 V_q \int  d^3 p \frac{\partial f}{\partial t}  \ln f \,,
\end{eqnarray}
where again $\int d^3 r =a^3 \int d^3 q =a^3 V_q$ and $N$ is the number of particles inside $V_q$. Using Eq.~\eqref{eq:boltzmann2}, the integral which appears in the last term of Eq.~\eqref{eq:dotH} may be written as
\be
 I  = \int  d^3 p  \frac{\partial f}{\partial t} \ln f= I_1 + I_2 \,,
\ee
where
\begin{eqnarray}
I_1&=& 3 (a^3 V_q)^{-1} \frac{d \ln \mathcal F}{dt} H\\
I_2&=& \int  d^3 p  \left(\frac{\partial  f}{\partial p^i}\frac{d \ln \left(a\mathcal F\right)}{dt}p^i\right) \ln f \,.
\end{eqnarray}
Integrating $I_2$ by parts one obtains
\begin{eqnarray}
\label{eq:integralI}
I_2 &=& - \int  d^3 p  f\frac{\partial }{\partial p^i} \left[\ln f\frac{d \ln \left(a \mathcal F\right)}{dt}p^i\right]\nonumber\\
&=&-3 (a^3 V_q)^{-1} \frac{d \ln \left(a \mathcal F\right)}{dt} H - \int  d^3 p  \frac{\partial f}{\partial p^i} \frac{d \ln \left(a \mathcal F\right)}{dt}p^i \nonumber \\
&=& 3 (a^3 V_q)^{-1} \frac{d \ln \left(a \mathcal F\right)}{dt} (N-H) \,.
\end{eqnarray}
Summing the various contributions, Eq.~\eqref{eq:dotH} finally becomes 
\begin{equation}
\label{eq:dotH1}
\frac{d H}{dt} = 3 \frac{d\ln \mathcal F}{dt} N \,.
\end{equation}

In general relativity $\mathcal F$ is equal to unity and, therefore, Boltzmann's $H$ is a constant in the absence of particle collisions. However, Eq. ~\eqref{eq:dotH1} implies that this is no longer true in the context of NMC theories of gravity. In this case, the evolution of Boltzmann's $H$ is directly coupled to the evolution of the universe. Boltzmann's $H$ may either increase or decrease, depending on whether $\mathcal F$ is a growing or a decaying function of time, respectively. This provides an explicitly demonstration that Boltzmann's $H$-theorem --- which states that $dH/dt \le 0$ --- may not hold in the context of NMC theories of gravity.

\subsection{An alternative derivation}

Here we provide an alternative derivation of Eq.~\eqref{eq:dotH1} (one of the main results of this paper). Consider two instants of time $t_A$ and $t_B$, with $a_A=1$. According to Eq.~\eqref{eq:3-force}, in the absence of collisions, $\vec p  \propto (a \mathcal F)^{-1}$. Therefore, assuming that the number of particles is conserved, Eq.~\eqref{V6} implies that
\begin{equation}
\frac{f_B}{f_A} \equiv \frac{f\left(t_B, \mathcal F_A \, \vec p /(a_B \mathcal F_B)  \right)}{f\left(t_A,\vec p\, \right)}=\frac{\mathcal V_{\rm 6A}}{\mathcal V_{\rm 6B}}=\left(\frac{\mathcal F_B}{\mathcal F_A}\right)^{3}\,.
\end{equation}
Hence,
\begin{eqnarray}
H_B&=&  a_B^3 V_q  \int  d^3 p_B  f_B  \ln  f_B \nonumber\\
&=& a_B^3 V_q  \int  d^3 p_A  (\mathcal F_A  /(a_B \mathcal F_B))^3   f_B  \ln  f_B \nonumber \\
&=&  V_q \int d^3 p_A f_A\ln \left(f_A ({\mathcal F}_B/{\mathcal F_A})^3\right)\nonumber\\
&=&V_q \int d^3 p_A f_A \ln f_A +  3  \ln\left({\mathcal F}_B/{\mathcal F}_A\right)      V_q \int d^3 p_A f_A \nonumber\\
&=&H_A + 3 \left( \ln({\mathcal F}_B)-  \ln ( {\mathcal F}_A)\right) N \,. \label{H2}
\end{eqnarray}
If $t_A$ and $t_B$ are sufficiently close, one can write $t_B-t_A=dt$,  $H_B-H_A = dH$, and $\ln(\mathcal F_B)-\ln(\mathcal F_A) = d \ln(\mathcal F)$. Then, diving Eq.~\eqref{H2} by $dt$ one obtains Eq.~\eqref{eq:dotH1}. This alternative derivation shows, perhaps even more explicitly, how the growth or decay of the magnitude of the linear momentum of the particles associated to the NMC  to gravity may contribute, respectively, to a decrease or an increase of Boltzmann's $H$. 

\section{Entropy}
\label{sec:entropy}

Consider a fluid of $N$ distinguishable point particles with Gibbs' and Boltzmann's entropies given respectively by
\begin{eqnarray}
S_{G}&=&-\int P_{N} \ln P_{N} d^{3} r_{1} d^{3} p_{1} \cdots d^{3} r_{N} d^{3} p_{N}\, \\
S_{B}&=&-N \int P \ln P d^{3} r d^{3} p\,.
\end{eqnarray}
where $P_{N}\left({\vec r}_1, {\vec p}_1, \ldots, {\vec r}_N, {\vec p}_N,t\right)$ and $P\left({\vec r}, {\vec p},t\right)$ are, respectively, the N-particle probability density function in $6N$-dimensional phase space and the single particle probability density function in 6-dimensional phase space. $P$ and $P_N$ are related by
\begin{equation}
P\left({\vec r}, {\vec p},t\right)=\int P_N d^{3} r_{2} d^{3} p_{2} \cdots d^{3} r_{N} d^{3} p_{N}
\end{equation}
These two definitions of the entropy have been shown to coincide only if
\begin{equation}
P_{N}\left({\vec r}_1, {\vec p}_1, \ldots, {\vec r}_N, {\vec p}_N,t\right)= \prod_{i=1}^N P\left({\vec r}_i, {\vec p}_i,t\right)\,,
\end{equation}
or, equivalently, if particle correlations can be neglected, as happens for an ideal gas \cite{1965AmJPh..33..391J}. In this case $S \equiv S_G=S_B$ (otherwise $S_G<S_B$ \cite{1965AmJPh..33..391J}). 

Consider a fixed comoving volume $V_q$ containing $N$ particles. Close to equilibrium $f\left({\vec r}, {\vec p},t\right)=NP\left({\vec r}, {\vec p},t\right)$  holds to an excellent approximation and, therefore
\begin{equation}
H=-S_B + N \ln N\,.
\end{equation}
Again, assuming that the particle number $N$ is fixed, Eq. ~\eqref{eq:dotH1} implies that
\begin{equation}
\label{eq:dotS}
\frac{d S_B}{dt}=-\frac{d H}{dt}  = - 3 \frac{d \ln \mathcal F}{dt} N\,.
\end{equation}
Hence, the entropy $S$ in a FLRW homogeneous and isotropic universe may decrease with cosmic time, as long as ${\mathcal F}$ grows with time. This shows that the second law of thermodynamics does not generally hold in the context of modified theories of gravity with a NMC between the gravitational and the matter fields (see also \cite{Azevedo:2019oah,Azevedo:2019krx}). Note that the entropy can never become negative since the number of states is always larger than or equal to unity --- in the $S \to 0$ limit all particles would condensate into a macroscopic zero-momentum state (in this limit the constant $N$ point particle assumption would no longer be valid).

\subsection{The collision term}

Adding a two-particle elastic scattering term to Eq.~\eqref{eq:continuity} results, under the assumption of molecular chaos, in a non-negative contribution to the entropy increase with cosmic time ---  this contribution vanishes for systems in thermodynamic equilibrium. This result holds independently of the NMC coupling to gravity, as acknowledged in \cite{Bertolami:2020ldj} where the standard calculation of the impact of the collision term has been performed without taking into account the momentum-dependent forces on the particles due to the NMC to gravity. However, as demonstrated in Secs.  \ref{sec:force} and \ref{sec:boltzmann}, these  momentum-dependent forces may be associated to a further decrease of the magnitude of the linear momentum of the particles (if $\mathcal F$ grows with time) contributing to the growth of Boltzmann's  $H$ (or, equivalently, to a decrease of the entropy). The existence of particle collisions, although extremely relevant in most cases, does not change this conclusion.

If the particles are non-relativistic, and assuming  thermodynamic equilibrium, $f(\vec p,t)$ follows a Maxwell-Boltzmann distribution. In a FLRW homogeneous and isotropic universe with a NMC to gravity the non-relativistic equilibrium distribution is maintained even if particle collisions are switched off at some later time, since the velocity of the individual particles would simply evolve as ${\vec v} \propto (a \mathcal F)^{-1}$ in the absence of collisions (see Eq.~\eqref{eq:3-force}) --- the temperature, in the case of non-relativistic particles, would evolve as $\mathcal T \propto v^2 \propto  (a \mathcal F)^{-2}$. 

On the other hand, in the case of indistinguishable relativistic particles with a  negligible chemical potential in thermodynamic equilibrium ${\mathcal T} \propto a^{-1} \mathcal F^{-1/4}$, so that the particle number density $n$ and the entropy density $s$ evolve as $n(\mathcal T) \propto s(\mathcal T) \propto {\mathcal T}^3 \propto a^{-3} \mathcal F^{-3/4}$. This implies that both the number of particles $N$ and the entropy $S$ in a fixed comoving volume are not conserved --- they evolve as $N \propto S \propto n a^3 \propto {\mathcal F}^{-3/4}$ (see \cite{Azevedo:2018nvi,Azevedo:2019oah} for more details). 

Unless ${\mathcal F}$ is a constant, the equilibrium distribution of the photons cannot be maintained after the Universe becomes transparent at a redshift $z \sim 10^3$, given that the number of photons of the cosmic background radiation is essentially conserved after that. Hence, a direct identification of Boltzmann's $H$ with  the entropy should not be made in this case. The requirement that the resulting spectral distortions be compatible with observations has been used to put stringent limits on the evolution of $\mathcal F$ after recombination \cite{Avelino:2018rsb}.

\subsection{The strength of gravity}
\label{sec:gstrength}

Existing cosmic microwave background and primordial nucleosynthesis constraints restrict the NMC theory of gravity studied in the present paper (or its most obvious generalization) to be very close to General Relativity ($\mathcal F=1$) at late times \cite{Avelino:2018rsb,Azevedo:2018nvi}. Prior to big bang nucleosynthesis the dynamics of $\mathcal F$ is much less constrained on observational grounds, but it is reasonable to expect that the cosmological principle and the existence of stable particles --- assumed throughout this work --- would still hold (at least after primordial inflation). This requires the avoidance of pathological instabilities, such as the Dolgov-Kawasaki instability, which, in the case of the action given in Eq.~\eqref{eq:action}, implies that $\mathcal F'' \mathcal{L}_{\rm m} \ge 0$ \cite{Faraoni:2007sn,Bertolami:2009cd}. 

Consider a scenario, free from pathological instabilities, in which the function $\mathcal F$ was much larger at early times than at late times (here, early and late refer to times much before and after primordial nucleosynthesis, respectively). In this scenario, the present value of Newton's gravitational constant is the result of a dynamical process associated to the decrease of $\mathcal F$, perhaps by many orders of magnitude, from early to late times. More importantly, the high entropy of the Universe and the weakness of gravity would be interrelated in this scenario.

\section{Conclusions}\label{sec:conc}

In this paper we have demonstrated that a violation of Boltzmann's $H$-theorem may occur in a homogeneous and isotropic FLRW universe whenever the NMC coupling function $\mathcal F$ is a growing function of the cosmic time. In order to show this, we started by computing the 4-acceleration of the point particles associated to the NMC to gravity. We then used the phase space continuity equation, expressing particle number conservation in six-dimensional phase space in the absence of collisions, to derive an expression for the evolution of Boltzmann's $H$ as a function of $\mathcal F$ (we also provided an alternative derivation of this relation). We have demonstrated that Boltzmann's $H$ may either increase or decrease, depending on whether the NMC coupling function $\mathcal F$ is a growing or a decaying function of cosmic time, respectively. We have considered the implications of this result for the evolution of the entropy of the matter fields, briefly discussing the role of collisions between the particles. Finally, we have highlighted the connection between the entropy of the universe and the strength of gravity in theories with a nonminimal matter-geometry coupling, showing, in particular, that the high entropy of the Universe and the weakness of gravity at the present time may be interrelated.

\begin{acknowledgments}

We thank Vasco Ferreira for fruitful discussions and for spotting a typo in one of our equations.
P.P.A. acknowledges the support from Fundação para a Ciência e a Tecnologia (FCT) through the Sabbatical Grant No. SFRH/BSAB/150322/2019. 	R.P.L.A. was supported by the Funda{\c c}\~ao para a Ci\^encia e Tecnologia (FCT, Portugal) grant SFRH/BD/132546/2017. Funding of this work has also been provided by FCT through national funds (PTDC/FIS-PAR/31938/2017) and by FEDER—Fundo Europeu de Desenvolvimento Regional through COMPETE2020 - Programa Operacional Competitividade e Internacionaliza{\c c}\~ao (POCI-01-0145-FEDER-031938), and through the research grants UID/FIS/04434/2019, UIDB/04434/2020 and UIDP/04434/2020.
\end{acknowledgments}

\bibliography{boltzmann}

\begin{thebibliography}{29}
\expandafter\ifx\csname natexlab\endcsname\relax\def\natexlab#1{#1}\fi
\expandafter\ifx\csname bibnamefont\endcsname\relax
  \def\bibnamefont#1{#1}\fi
\expandafter\ifx\csname bibfnamefont\endcsname\relax
  \def\bibfnamefont#1{#1}\fi
\expandafter\ifx\csname citenamefont\endcsname\relax
  \def\citenamefont#1{#1}\fi
\expandafter\ifx\csname url\endcsname\relax
  \def\url#1{\texttt{#1}}\fi
\expandafter\ifx\csname urlprefix\endcsname\relax\def\urlprefix{URL }\fi
\providecommand{\bibinfo}[2]{#2}
\providecommand{\eprint}[2][]{\url{#2}}

\bibitem[{\citenamefont{Will}(2014)}]{Will:2014kxa}
\bibinfo{author}{\bibfnamefont{C.~M.} \bibnamefont{Will}},
  \bibinfo{journal}{Living Rev. Rel.} \textbf{\bibinfo{volume}{17}},
  \bibinfo{pages}{4} (\bibinfo{year}{2014}), \eprint{1403.7377}.

\bibitem[{\citenamefont{Riess et~al.}(1998)}]{Riess:1998cb}
\bibinfo{author}{\bibfnamefont{A.~G.} \bibnamefont{Riess}} \bibnamefont{et~al.}
  (\bibinfo{collaboration}{Supernova Search Team}), \bibinfo{journal}{Astron.
  J.} \textbf{\bibinfo{volume}{116}}, \bibinfo{pages}{1009}
  (\bibinfo{year}{1998}), \eprint{astro-ph/9805201}.

\bibitem[{\citenamefont{Perlmutter et~al.}(1999)}]{Perlmutter:1998np}
\bibinfo{author}{\bibfnamefont{S.}~\bibnamefont{Perlmutter}}
  \bibnamefont{et~al.} (\bibinfo{collaboration}{Supernova Cosmology Project}),
  \bibinfo{journal}{Astrophys. J.} \textbf{\bibinfo{volume}{517}},
  \bibinfo{pages}{565} (\bibinfo{year}{1999}), \eprint{astro-ph/9812133}.

\bibitem[{\citenamefont{Aghanim et~al.}(2018)}]{Aghanim:2018eyx}
\bibinfo{author}{\bibfnamefont{N.}~\bibnamefont{Aghanim}} \bibnamefont{et~al.}
  (\bibinfo{collaboration}{Planck}) (\bibinfo{year}{2018}),
  \eprint{1807.06209}.

\bibitem[{\citenamefont{Akrami et~al.}(2018)}]{Akrami:2018odb}
\bibinfo{author}{\bibfnamefont{Y.}~\bibnamefont{Akrami}} \bibnamefont{et~al.}
  (\bibinfo{collaboration}{Planck}) (\bibinfo{year}{2018}),
  \eprint{1807.06211}.

\bibitem[{\citenamefont{Nojiri and Odintsov}(2004)}]{Nojiri:2004bi}
\bibinfo{author}{\bibfnamefont{S.}~\bibnamefont{Nojiri}} \bibnamefont{and}
  \bibinfo{author}{\bibfnamefont{S.~D.} \bibnamefont{Odintsov}},
  \bibinfo{journal}{Phys. Lett.} \textbf{\bibinfo{volume}{B599}},
  \bibinfo{pages}{137} (\bibinfo{year}{2004}), \eprint{astro-ph/0403622}.

\bibitem[{\citenamefont{Bouhmadi-Lopez and Wands}(2005)}]{BouhmadiLopez:2004ys}
\bibinfo{author}{\bibfnamefont{M.}~\bibnamefont{Bouhmadi-Lopez}}
  \bibnamefont{and} \bibinfo{author}{\bibfnamefont{D.}~\bibnamefont{Wands}},
  \bibinfo{journal}{Phys. Rev.} \textbf{\bibinfo{volume}{D71}},
  \bibinfo{pages}{024010} (\bibinfo{year}{2005}), \eprint{hep-th/0408061}.

\bibitem[{\citenamefont{Allemandi et~al.}(2005)\citenamefont{Allemandi,
  Borowiec, Francaviglia, and Odintsov}}]{Allemandi:2005qs}
\bibinfo{author}{\bibfnamefont{G.}~\bibnamefont{Allemandi}},
  \bibinfo{author}{\bibfnamefont{A.}~\bibnamefont{Borowiec}},
  \bibinfo{author}{\bibfnamefont{M.}~\bibnamefont{Francaviglia}},
  \bibnamefont{and} \bibinfo{author}{\bibfnamefont{S.~D.}
  \bibnamefont{Odintsov}}, \bibinfo{journal}{Phys. Rev.}
  \textbf{\bibinfo{volume}{D72}}, \bibinfo{pages}{063505}
  (\bibinfo{year}{2005}), \eprint{gr-qc/0504057}.

\bibitem[{\citenamefont{Bertolami et~al.}(2007)\citenamefont{Bertolami,
  Boehmer, Harko, and Lobo}}]{Bertolami:2007gv}
\bibinfo{author}{\bibfnamefont{O.}~\bibnamefont{Bertolami}},
  \bibinfo{author}{\bibfnamefont{C.~G.} \bibnamefont{Boehmer}},
  \bibinfo{author}{\bibfnamefont{T.}~\bibnamefont{Harko}}, \bibnamefont{and}
  \bibinfo{author}{\bibfnamefont{F.~S.~N.} \bibnamefont{Lobo}},
  \bibinfo{journal}{Phys. Rev. D} \textbf{\bibinfo{volume}{75}},
  \bibinfo{pages}{104016} (\bibinfo{year}{2007}), \eprint{0704.1733}.

\bibitem[{\citenamefont{Sotiriou and Faraoni}(2008)}]{Sotiriou:2008it}
\bibinfo{author}{\bibfnamefont{T.~P.} \bibnamefont{Sotiriou}} \bibnamefont{and}
  \bibinfo{author}{\bibfnamefont{V.}~\bibnamefont{Faraoni}},
  \bibinfo{journal}{Class. Quant. Grav.} \textbf{\bibinfo{volume}{25}},
  \bibinfo{pages}{205002} (\bibinfo{year}{2008}), \eprint{0805.1249}.

\bibitem[{\citenamefont{Bertolami et~al.}(2008)\citenamefont{Bertolami, Lobo,
  and Paramos}}]{Bertolami:2008ab}
\bibinfo{author}{\bibfnamefont{O.}~\bibnamefont{Bertolami}},
  \bibinfo{author}{\bibfnamefont{F.~S.} \bibnamefont{Lobo}}, \bibnamefont{and}
  \bibinfo{author}{\bibfnamefont{J.}~\bibnamefont{Paramos}},
  \bibinfo{journal}{Phys. Rev. D} \textbf{\bibinfo{volume}{78}},
  \bibinfo{pages}{064036} (\bibinfo{year}{2008}), \eprint{0806.4434}.

\bibitem[{\citenamefont{Harko and Lobo}(2010)}]{Harko:2010mv}
\bibinfo{author}{\bibfnamefont{T.}~\bibnamefont{Harko}} \bibnamefont{and}
  \bibinfo{author}{\bibfnamefont{F.~S.~N.} \bibnamefont{Lobo}},
  \bibinfo{journal}{Eur. Phys. J.} \textbf{\bibinfo{volume}{C70}},
  \bibinfo{pages}{373} (\bibinfo{year}{2010}), \eprint{1008.4193}.

\bibitem[{\citenamefont{Harko et~al.}(2011)\citenamefont{Harko, Lobo, Nojiri,
  and Odintsov}}]{Harko:2011kv}
\bibinfo{author}{\bibfnamefont{T.}~\bibnamefont{Harko}},
  \bibinfo{author}{\bibfnamefont{F.~S.~N.} \bibnamefont{Lobo}},
  \bibinfo{author}{\bibfnamefont{S.}~\bibnamefont{Nojiri}}, \bibnamefont{and}
  \bibinfo{author}{\bibfnamefont{S.~D.} \bibnamefont{Odintsov}},
  \bibinfo{journal}{Phys. Rev.} \textbf{\bibinfo{volume}{D84}},
  \bibinfo{pages}{024020} (\bibinfo{year}{2011}), \eprint{1104.2669}.

\bibitem[{\citenamefont{Harko et~al.}(2013)\citenamefont{Harko, Lobo, and
  Minazzoli}}]{Harko:2012hm}
\bibinfo{author}{\bibfnamefont{T.}~\bibnamefont{Harko}},
  \bibinfo{author}{\bibfnamefont{F.~S.~N.} \bibnamefont{Lobo}},
  \bibnamefont{and}
  \bibinfo{author}{\bibfnamefont{O.}~\bibnamefont{Minazzoli}},
  \bibinfo{journal}{Phys. Rev.} \textbf{\bibinfo{volume}{D87}},
  \bibinfo{pages}{047501} (\bibinfo{year}{2013}), \eprint{1210.4218}.

\bibitem[{\citenamefont{Avelino and Sousa}(2018)}]{Avelino:2018qgt}
\bibinfo{author}{\bibfnamefont{P.~P.} \bibnamefont{Avelino}} \bibnamefont{and}
  \bibinfo{author}{\bibfnamefont{L.}~\bibnamefont{Sousa}},
  \bibinfo{journal}{Phys. Rev.} \textbf{\bibinfo{volume}{D97}},
  \bibinfo{pages}{064019} (\bibinfo{year}{2018}), \eprint{1802.03961}.

\bibitem[{\citenamefont{Avelino and Azevedo}(2018)}]{Avelino:2018rsb}
\bibinfo{author}{\bibfnamefont{P.~P.} \bibnamefont{Avelino}} \bibnamefont{and}
  \bibinfo{author}{\bibfnamefont{R.~P.~L.} \bibnamefont{Azevedo}},
  \bibinfo{journal}{Phys. Rev.} \textbf{\bibinfo{volume}{D97}},
  \bibinfo{pages}{064018} (\bibinfo{year}{2018}), \eprint{1802.04760}.

\bibitem[{\citenamefont{Azevedo and Avelino}(2018)}]{Azevedo:2018nvi}
\bibinfo{author}{\bibfnamefont{R.~P.~L.} \bibnamefont{Azevedo}}
  \bibnamefont{and} \bibinfo{author}{\bibfnamefont{P.~P.}
  \bibnamefont{Avelino}}, \bibinfo{journal}{Phys. Rev.}
  \textbf{\bibinfo{volume}{D98}}, \bibinfo{pages}{064045}
  (\bibinfo{year}{2018}), \eprint{1807.00798}.

\bibitem[{\citenamefont{{Jaynes}}(1965)}]{1965AmJPh..33..391J}
\bibinfo{author}{\bibfnamefont{E.~T.} \bibnamefont{{Jaynes}}},
  \bibinfo{journal}{American Journal of Physics} \textbf{\bibinfo{volume}{33}},
  \bibinfo{pages}{391} (\bibinfo{year}{1965}).

\bibitem[{\citenamefont{Bertolami and Gomes}(2020)}]{Bertolami:2020ldj}
\bibinfo{author}{\bibfnamefont{O.}~\bibnamefont{Bertolami}} \bibnamefont{and}
  \bibinfo{author}{\bibfnamefont{C.}~\bibnamefont{Gomes}}
  (\bibinfo{year}{2020}), \eprint{2002.08184}.

\bibitem[{\citenamefont{Polyakov and Schweitzer}(2018)}]{Polyakov:2018zvc}
\bibinfo{author}{\bibfnamefont{M.~V.} \bibnamefont{Polyakov}} \bibnamefont{and}
  \bibinfo{author}{\bibfnamefont{P.}~\bibnamefont{Schweitzer}},
  \bibinfo{journal}{Int. J. Mod. Phys.} \textbf{\bibinfo{volume}{A33}},
  \bibinfo{pages}{1830025} (\bibinfo{year}{2018}), \eprint{1805.06596}.

\bibitem[{\citenamefont{Avelino}(2019)}]{Avelino:2019esh}
\bibinfo{author}{\bibfnamefont{P.~P.} \bibnamefont{Avelino}},
  \bibinfo{journal}{Phys. Lett.} \textbf{\bibinfo{volume}{B795}},
  \bibinfo{pages}{627} (\bibinfo{year}{2019}), \eprint{1902.01318}.

\bibitem[{\citenamefont{Ayaita et~al.}(2012)\citenamefont{Ayaita, Weber, and
  Wetterich}}]{Ayaita:2011ay}
\bibinfo{author}{\bibfnamefont{Y.}~\bibnamefont{Ayaita}},
  \bibinfo{author}{\bibfnamefont{M.}~\bibnamefont{Weber}}, \bibnamefont{and}
  \bibinfo{author}{\bibfnamefont{C.}~\bibnamefont{Wetterich}},
  \bibinfo{journal}{Phys. Rev.} \textbf{\bibinfo{volume}{D85}},
  \bibinfo{pages}{123010} (\bibinfo{year}{2012}), \eprint{1112.4762}.

\bibitem[{\citenamefont{Ferreira et~al.}(2020)\citenamefont{Ferreira, Avelino,
  and Azevedo}}]{Ferreira:2020fma}
\bibinfo{author}{\bibfnamefont{V.~M.~C.} \bibnamefont{Ferreira}},
  \bibinfo{author}{\bibfnamefont{P.~P.} \bibnamefont{Avelino}},
  \bibnamefont{and} \bibinfo{author}{\bibfnamefont{R.~P.~L.}
  \bibnamefont{Azevedo}} (\bibinfo{year}{2020}), \eprint{2005.07739}.

\bibitem[{\citenamefont{Brown}(1993)}]{Brown:1992kc}
\bibinfo{author}{\bibfnamefont{J.~D.} \bibnamefont{Brown}},
  \bibinfo{journal}{Class. Quant. Grav.} \textbf{\bibinfo{volume}{10}},
  \bibinfo{pages}{1579} (\bibinfo{year}{1993}), \eprint{gr-qc/9304026}.

\bibitem[{\citenamefont{Schutz and Sorkin}(1977)}]{SCHUTZ19771}
\bibinfo{author}{\bibfnamefont{B.~F.} \bibnamefont{Schutz}} \bibnamefont{and}
  \bibinfo{author}{\bibfnamefont{R.}~\bibnamefont{Sorkin}},
  \bibinfo{journal}{Annals of Physics} \textbf{\bibinfo{volume}{107}},
  \bibinfo{pages}{1 } (\bibinfo{year}{1977}), ISSN \bibinfo{issn}{0003-4916}.

\bibitem[{\citenamefont{Azevedo and
  Avelino}(2019{\natexlab{a}})}]{Azevedo:2019oah}
\bibinfo{author}{\bibfnamefont{R.~P.~L.} \bibnamefont{Azevedo}}
  \bibnamefont{and} \bibinfo{author}{\bibfnamefont{P.~P.}
  \bibnamefont{Avelino}} (\bibinfo{year}{2019}{\natexlab{a}}),
  \eprint{1908.02629}.

\bibitem[{\citenamefont{Azevedo and
  Avelino}(2019{\natexlab{b}})}]{Azevedo:2019krx}
\bibinfo{author}{\bibfnamefont{R.~P.~L.} \bibnamefont{Azevedo}}
  \bibnamefont{and} \bibinfo{author}{\bibfnamefont{P.~P.}
  \bibnamefont{Avelino}}, \bibinfo{journal}{Phys. Rev.}
  \textbf{\bibinfo{volume}{D99}}, \bibinfo{pages}{064027}
  (\bibinfo{year}{2019}{\natexlab{b}}), \eprint{1901.06299}.

\bibitem[{\citenamefont{Faraoni}(2007)}]{Faraoni:2007sn}
\bibinfo{author}{\bibfnamefont{V.}~\bibnamefont{Faraoni}},
  \bibinfo{journal}{Phys. Rev. D} \textbf{\bibinfo{volume}{76}},
  \bibinfo{pages}{127501} (\bibinfo{year}{2007}), \eprint{0710.1291}.

\bibitem[{\citenamefont{Bertolami and Sequeira}(2009)}]{Bertolami:2009cd}
\bibinfo{author}{\bibfnamefont{O.}~\bibnamefont{Bertolami}} \bibnamefont{and}
  \bibinfo{author}{\bibfnamefont{M.~C.} \bibnamefont{Sequeira}},
  \bibinfo{journal}{Phys. Rev. D} \textbf{\bibinfo{volume}{79}},
  \bibinfo{pages}{104010} (\bibinfo{year}{2009}), \eprint{0903.4540}.

\end{thebibliography}
 	
 \end{document}